\documentclass{article}

\usepackage{PRIMEarxiv}

\usepackage[utf8]{inputenc} 
\usepackage[T1]{fontenc}    
\usepackage{hyperref}       
\usepackage{url}            
\usepackage{booktabs}       
\usepackage{amsfonts}       
\usepackage{nicefrac}       
\usepackage{microtype}      
\usepackage{fancyhdr}       
\usepackage{graphicx}       
\graphicspath{{media/}}     
\usepackage{amsmath}        
\usepackage{float} 

\title{Feature Selection via GANs (GANFS): Enhancing Machine Learning Models for DDoS Mitigation
}

\author{
  Harsh Patel \\
  School of Electrical and Computer Science \\
  University of Ottawa \\
  Ottawa, Canada\\
  \texttt{hpate033@uottawa.ca} \\ %
}

\begin{document}
\maketitle

\begin{abstract} %
Distributed Denial of Service (DDoS) attacks represent a persistent and evolving threat to modern networked systems, capable of causing large-scale service disruptions. The complexity of such attacks, often hidden within high-dimensional and redundant network traffic data, necessitates robust and intelligent feature selection techniques for effective detection. Traditional methods such as filter-based, wrapper-based, and embedded approaches, each offer strengths but struggle with scalability or adaptability in complex attack environments. In this study, we explore these existing techniques through a detailed comparative analysis and highlight their limitations when applied to large-scale DDoS detection tasks. Building upon these insights, we introduce a novel Generative Adversarial Network-based Feature Selection (GANFS) method that leverages adversarial learning dynamics to identify the most informative features. By training a GAN exclusively on attack traffic and employing a perturbation-based sensitivity analysis on the Discriminator, GANFS effectively ranks feature importance without relying on full supervision. Experimental evaluations using the CIC-DDoS2019 dataset demonstrate that GANFS not only improves the accuracy of downstream classifiers but also enhances computational efficiency by significantly reducing feature dimensionality. These results point to the potential of integrating generative learning models into cybersecurity pipelines to build more adaptive and scalable detection systems.
\end{abstract}

\keywords{DDoS \and Feature Selection \and GAN \and Machine Learning \and Artificial Intelligence \and Cybersecurity \and Network Security}

\section{Introduction} %
In the digital age, the reliability and availability of networked services are foundational to both personal communication and enterprise operations. However, these services remain highly vulnerable to Distributed Denial of Service (DDoS) attacks, which are among the most disruptive forms of cyber threats. By overwhelming servers, routers, or application services with massive volumes of malicious traffic, DDoS attacks render legitimate access impossible, causing significant financial and reputational damage. With the increasing complexity of network infrastructures and the proliferation of connected devices, including those in the Internet of Things (IoT), the scale and sophistication of DDoS attacks continue to rise \cite{kurniawan2022comparison, gaur2021analysis}.

The problem addressed in this study is the challenge of identifying and selecting the most relevant features from high-dimensional network traffic data to detect DDoS attacks effectively. Network flow datasets, such as CIC-DDoS2019, typically contain upwards of 80 features derived from raw packet streams. While rich in information, these datasets include redundant, irrelevant, or noisy features that hinder machine learning (ML) model performance. Feeding such high-dimensional data into classifiers can result in increased training times, higher memory consumption, and reduced detection accuracy due to overfitting \cite{wang2022curse, maslan2020feature}. Solving this problem is critical for enabling scalable, efficient, and accurate DDoS detection systems. Effective feature selection (FS) can significantly reduce dimensionality, enhance interpretability, and improve the real-time capabilities of intrusion detection systems (IDS). It is especially important in resource-constrained environments where quick decision-making and lightweight models are essential. For instance, in edge-based or IoT security frameworks.

Several approaches have been explored to tackle this issue, and each belongs to one of three broad categories: filter-based, wrapper-based, and embedded methods \cite{kurniawan2022comparison, wang2022curse, subasri2022machine}. Filter-based methods, such as Mutual Information (MI), Chi-Square, and Information Gain (IG), assess the statistical relationship between each feature and the target variable independently of any ML model \cite{wang2022curse, maslan2020feature}. These techniques are computationally efficient and scale well to large datasets, but they often ignore interactions between features, leading to the selection of suboptimal subsets. Wrapper-based methods, including Recursive Feature Elimination (RFE) and Genetic Algorithms (GA), evaluate different feature subsets by training models and selecting those with the best performance. Although typically more accurate, these methods are computationally expensive and impractical for real-time or large-scale applications \cite{aamir2019ddos}. Embedded methods, such as Lasso (L1 regularization), Random Forests (RF), and XGBoost, integrate feature selection within the model training process. These offer a balance between efficiency and accuracy, but often their outputs are highly model-specific and lack generalizability across different classifiers \cite{kurniawan2022comparison, subasri2022machine, gaur2022fsmdad}.

Despite the progress in traditional FS techniques, a key limitation remains: most approaches fail to capture complex, non-linear feature interactions or adapt well to dynamic and adversarial environments. They also typically require full supervision, with balanced and labelled datasets, a condition that is often unrealistic in cybersecurity settings. Moreover, many conventional methods do not leverage the potential of deep learning, especially generative modelling, which can learn intricate data distributions without relying on strict labelling or pre-defined relationships.

To address these gaps, this study introduces a novel feature selection method called Generative Adversarial Network-based Feature Selection (GANFS). GANFS is designed to exploit the generative-discriminative dynamics of GANs to identify features that are most essential for characterizing DDoS attacks. The algorithm operates by first training a GAN using only DDoS attack traffic (i.e., samples labelled as malicious). The Generator attempts to produce synthetic attack samples that mimic real attack traffic, while the Discriminator learns to distinguish real from generated samples. Once trained, the Generator is discarded, and the Discriminator is used to perform sensitivity analysis through feature perturbation. The rationale is that features causing the largest changes in the Discriminator's output confidence are the most important for defining true attack behavior.

We selected the GANFS algorithm for three primary reasons:
\begin{itemize}
    \item GANs are uniquely suited for modelling high-dimensional and non-linear data distributions, which aligns well with the complex nature of network traffic.
    \item The adversarial learning setup naturally highlights features that are “difficult to fake,” thus indirectly revealing those most critical for distinguishing attacks.
    \item GANFS eliminates the need for repeated model retraining as in wrapper methods, making it more scalable and practical for large datasets.
\end{itemize}
Our choice was also informed by the limitations observed in prior FS techniques. Filter-based methods like MI and Chi-Square quickly identified features with high statistical relevance but failed to generalize well in downstream classifiers, likely due to ignoring inter-feature dependencies \cite{wang2022curse, maslan2020feature}. Wrapper-based methods such as RFE yielded higher classification accuracy but at significant computational cost, making them unsuitable for real-time or large-scale analysis \cite{aamir2019ddos}. Embedded methods like Random Forest performed well in identifying some key features, such as packet counts and byte rates but were prone to instability across different runs and sensitive to imbalanced class distributions \cite{kurniawan2022comparison, subasri2022machine}.

The key innovation of GANFS lies in its unsupervised, model-agnostic feature evaluation strategy. By treating the Discriminator as a proxy for feature quality and measuring its sensitivity to perturbations in individual features, GANFS provides a fine-grained, data-driven ranking without requiring external classifiers or performance metrics. This allows for a lightweight, one-pass evaluation of feature importance after GAN training.

In brief, this research makes the following contributions:
\begin{itemize} %
    \item It presents a detailed empirical evaluation of existing feature selection techniques across multiple categories using the CIC-DDoS2019 dataset.
    \item It proposes GANFS, a novel, GAN-driven feature selection framework optimized for detecting adversarial traffic patterns.
    \item It introduces a robust sensitivity analysis method that quantifies feature importance based on Discriminator confidence variation.
    \item It demonstrates through experiments that GANFS outperforms traditional methods in selecting compact yet highly informative feature subsets, leading to improved classification accuracy and reduced computational cost.
\end{itemize}
By integrating generative modelling with sensitivity-based ranking, GANFS represents a significant step toward more intelligent, adaptive, and scalable feature selection in cybersecurity applications. The findings of this work not only enhance the state of the art in DDoS detection but also provide a foundation for applying adversarial learning to broader security-related data mining challenges.

\section{State of the art review} %
DDoS attacks exploit vulnerabilities across diverse environments, including cloud services, IoT devices, and traditional networks. Attackers often leverage botnets—large networks of compromised devices—to generate massive volumes of traffic targeting specific victims. The primary challenge lies in accurately differentiating attack traffic from legitimate user activity within high-dimensional datasets characterized by numerous, potentially irrelevant features.

\subsection{Need for feature selection in DDOS Detection} %
Raw network traffic data typically encompasses a wide array of features, such as packet rates, protocol flags, packet sizes, and inter-arrival times. However, many of these features may be redundant or irrelevant for distinguishing between normal and malicious traffic. This redundancy can lead to several problems:
\begin{itemize} %
    \item Increased Computational Costs: Processing high-dimensional data requires significant computational resources, slowing down the detection process and making real-time analysis difficult.
    \item Model Overfitting: Machine learning models that are trained on irrelevant features may exhibit overfitting of the training data, resulting in poor generalization and reduced accuracy on new, unseen data.
    \item Reduced Detection Accuracy: The presence of irrelevant features can obscure the patterns associated with DDoS attacks, making it harder for detection models to identify malicious traffic.
\end{itemize}
Feature selection addresses these challenges by recognizing the most informative features, decreasing the dimensionality of the data, and improving the performance of DDoS detection models. Effective FS techniques can significantly enhance detection accuracy, reduce computational overhead, and improve the ability to generalize across different attack types and network environments.

\subsection{Overview of Feature Selection Methods} %
Feature selection methods can be broadly categorized into four main approaches: filter-based methods, wrapper-based methods, embedded methods, and hybrid methods. In the context of DDoS detection, each of these methods has unique benefits and trade-offs regarding accuracy, generalizability, and computing efficiency.

\subsubsection{Filter-Based Methods} %
Filter methods select features based on statistical measures evaluated independently of any specific machine learning algorithm. These techniques are appropriate for big datasets since they are scalable and computationally efficient.
\paragraph{Mutual Information (MI)} %
Mutual Information (MI) quantifies the statistical dependence between input features and class labels. A higher MI score indicates a stronger relationship and suggests greater relevance for classification tasks. This measure is particularly effective for identifying features that contain valuable predictive information without relying on any model-specific assumptions. For instance, Abu Bakar et al. \cite{bakar2023intelligent} introduced an intelligent agent-based detection framework that integrated MI for automatic feature extraction and selection. Their system achieved a high detection accuracy of 99.7\% in an IoT-based DDoS detection scenario, showcasing the utility of MI in real-world applications. However, despite its efficiency, MI does not account for interactions between features and may lead to the selection of redundant or collinear features. This makes it less effective when dealing with complex datasets where meaningful patterns emerge only from combinations of attributes.

\paragraph{Chi-Square Test} %
The Chi-Square test evaluates the independence between categorical features and class labels, making it particularly suitable for identifying features that are highly correlated with specific target classes. It is a widely used statistical method in feature selection, especially when dealing with discrete or categorical data. Muhammad Aamir et al. \cite{aamir2019ddos} employed the Chi-Square test to reduce the number of features in a DDoS detection model by 68\%, with minimal loss in accuracy, demonstrating the test’s practical effectiveness. Gaur and Kumar \cite{gaur2021analysis, gaur2022fsmdad} also utilized Chi-Square alongside other filter techniques like ANOVA and Extra Trees across multiple classifiers, which further validated its reliability. While Chi-Square is straightforward and computationally efficient, it assumes feature independence and is sensitive to the distribution of data. This can result in misleading rankings in cases where feature interdependencies are important or when the feature space contains continuous variables that are poorly discretized.

\paragraph{Information Gain (IG)} %
Information Gain (IG) measures the reduction in entropy achieved by partitioning the dataset based on a particular feature. Features that contribute to a significant decrease in uncertainty are considered more informative and thus preferred for model training. This metric is commonly used in decision tree algorithms and filter-based FS pipelines. Subasri et al. \cite{subasri2022machine} applied Information Gain in combination with data reduction techniques for feature selection in Named Data Networking (NDN), yielding efficient and accurate detection of DDoS attacks. Despite its utility, IG tends to favour features with a large number of unique values, which can introduce bias and lead to overfitting. Furthermore, similar to other univariate filter methods, IG fails to capture complex dependencies between multiple features.

\paragraph{Correlation-Based Feature Selection} %
Correlation-based feature selection involves identifying and removing features that are highly correlated with each other to reduce redundancy and improve model interpretability. By focusing on reducing multicollinearity, this approach simplifies the input space and can enhance generalization in machine learning models. Azadeh Golduzian \cite{golduzian2023predict} utilized correlation analysis through visual heatmaps to pinpoint and exclude redundant attributes, leading to a more streamlined feature set and improved detection performance. While this method is effective in pruning excessive or duplicative data, it does not directly measure a feature’s relationship with the target variable and may inadvertently remove useful features if the correlation threshold is set too strictly. Moreover, it assumes linear relationships and may not detect non-linear dependencies that could be valuable in classification.

\subsubsection{Wrapper-Based Methods} %
Wrapper methods are a subclass of machine learning algorithms that are designed to evaluate feature subsets. These methods employ a specific machine learning model for training and evaluation, resulting in more accurate outcomes when compared to filter methods. However, it should be noted that wrapper methods are also more computationally expensive.

\paragraph{Recursive Feature Elimination (RFE)} %
Recursive Feature Elimination (RFE) is an iterative method that progressively removes the least important features according to model-based rankings. The process continues until an optimal subset is identified that yields high predictive accuracy. Aamir and Zaidi \cite{aamir2019ddos} successfully implemented RFE in their study on DDoS attack detection, achieving strong classification performance with a reduced feature set. By aligning feature selection with the specific learning algorithm used for classification, RFE offers high precision in identifying relevant subsets. However, the need to retrain the model multiple times makes RFE computationally expensive, especially when applied to large, high-dimensional datasets. This limits its scalability and real-time applicability.

\paragraph{Genetic Algorithms (GA)} %
Genetic Algorithms (GAs) take inspiration from evolutionary biology to explore the feature space. They treat each feature subset as an individual and apply operations such as selection, crossover, and mutation to evolve toward an optimal solution. This global search capability allows GAs to navigate complex, non-convex spaces that might trap greedy methods. However, while powerful, GAs are also resource-intensive and sensitive to parameter settings such as population size, mutation rate, and selection strategy. Without proper tuning and validation, they may converge to suboptimal solutions or require substantial computational time.

\paragraph{Sequential Feature Selection (SFS)} %
Sequential Feature Selection (SFS) methods add or remove features one at a time based on their marginal impact on model performance. Forward selection starts with an empty set and adds features, while backward elimination begins with all features and prunes them iteratively. Though simpler and more interpretable than stochastic methods like GAs, SFS is prone to local minima and can be inefficient for large feature spaces. Additionally, it assumes that feature contributions are independent and additive, which may not hold in complex cybersecurity datasets like those used in DDoS detection.

\subsubsection{Embedded Methods} %
Many classification models integrate feature selection as a part. Embedded methods use these models to perform feature selection. These methods often provide a good balance between accuracy and computational efficiency.

\paragraph{Lasso (L1 Regularization)} %
Lasso is a linear model that incorporates L1 regularization into the loss function, effectively penalizing the absolute values of the feature coefficients. This penalty induces sparsity by driving the coefficients of less informative features to zero, making Lasso a natural choice for embedded feature selection. It is particularly useful when dealing with high-dimensional data where interpretability is important. Studies such as Maslan et al. \cite{maslan2020feature} have demonstrated the effectiveness of Lasso in reducing overfitting in machine learning models applied to DDoS detection. However, Lasso assumes linearity and may not capture non-linear interactions between features. It can also struggle when features are highly correlated, often arbitrarily selecting one and ignoring the rest.

\paragraph{Random Forest Feature Importance} %
Random Forest (RF) models provide intrinsic feature importance scores by measuring how much each feature reduces impurity across all trees in the ensemble. The mean decrease in Gini impurity or mean decrease in accuracy is typically used to rank features. In the context of cybersecurity, Kurniawan et al. \cite{kurniawan2022comparison} applied RF to Software Defined Networks (SDN) and identified pktcount as a dominant feature for DDoS attack detection. Similar findings have been echoed in comparative analyses such as Gaur and Kumar \cite{gaur2021analysis}, where RF was found to consistently outperform other classifiers in detecting malicious traffic when trained on optimized feature subsets. While RF is robust and handles feature interactions well, the importance rankings may vary across different runs due to its stochastic nature, and it may overemphasize features with many distinct values.

\paragraph{Extreme Gradient Boosting (XGBoost)} %
XGBoost is an advanced gradient boosting algorithm that also provides built-in feature importance metrics. It calculates importance based on the frequency and quality of feature usage in tree splits across boosted iterations. In the work of Golduzian \cite{golduzian2023predict}, XGBoost was employed with a reduced feature set for DDoS detection and achieved notably high performance while maintaining low computational overhead. The algorithm is known for its scalability and effectiveness in imbalanced datasets, making it suitable for real-world network traffic analysis. However, like RF, its feature rankings are model-specific and can fluctuate depending on hyperparameter choices and dataset characteristics. It is also sensitive to class imbalance and noise in the data, which may skew the importance scores.

\subsubsection{Hybrid Methods} %
Hybrid methods combine multiple feature selection techniques to leverage their complementary strengths.

\paragraph{Combining Filter and Wrapper Methods} %
One such approach is seen in the Feature and Model Selection (FAMS) framework introduced by Ma et al. \cite{ma2023ddos}, which integrates multiple FS techniques with model selection to optimize detection performance. By combining filter, wrapper, and embedded strategies, the framework was able to generalize effectively across different types of DDoS attacks, yielding improved robustness and reliability. Similarly, Abu Bakar et al. \cite{bakar2023intelligent} proposed an intelligent agent-based system that automated the feature extraction and selection process. Their system utilized sequential feature selection along with domain-specific rules to refine the FS pipeline, resulting in superior performance in IoT environments.

\paragraph{Ensemble Feature Selection} %
Ensemble feature selection methods further extend this idea by applying multiple FS algorithms and aggregating their results—often through voting, ranking fusion, or statistical consensus—to achieve a more stable feature subset. This was illustrated in the work of Mishra et al. \cite{mishra2022defensive}, who combined feature selection with multi-classifier systems to improve robustness against different types of attacks. Although hybrid approaches often outperform individual techniques in both accuracy and generalizability, they also introduce significant complexity. Tuning and validating multiple algorithms simultaneously can be computationally intensive and increases the risk of overfitting, especially when applied without rigorous cross-validation.

\subsection{Existing Deep Learning Methods} %
In recent years, deep learning models have gained traction for cybersecurity tasks, including DDoS detection and network anomaly identification. Architectures such as Convolutional Neural Networks (CNNs), Autoencoders, Recurrent Neural Networks (RNNs), and Transformer-based models have been explored to process high-dimensional network traffic data and detect malicious behaviour. While each model type offers unique advantages, the choice of a Generative Adversarial Network (GAN) for feature selection in this study is motivated by both the problem’s nature and the specific demands of feature relevance extraction.

\subsubsection{Convolutional Neural Networks (CNNs)} %
CNNs are highly effective for pattern recognition tasks and have been applied to network traffic classification by treating flow-based features as image-like inputs \cite{raza2024feature, golduzian2023predict}. They are particularly good at detecting spatial hierarchies and local dependencies, which can be useful when features have positional structure. However, in feature selection tasks, CNNs offer limited interpretability and do not inherently provide a mechanism to rank input features. Feature relevance in CNNs typically requires additional post hoc explainability tools, which can be complex and indirect.

\subsubsection{Autoencoders} %
Autoencoders (AEs) and Variational Autoencoders (VAEs) are unsupervised models used for dimensionality reduction and anomaly detection. In DDoS contexts, AEs can learn compressed representations of normal traffic and flag deviations as anomalies \cite{gaur2022fsmdad}. While this makes AEs suitable for detecting outliers or reconstructing inputs, they are not optimized for distinguishing subtle differences between real and synthetic attack samples. Moreover, feature selection using AEs often depends on reconstruction error or learned latent space variance, which may not capture true discriminative power among features.

\subsubsection{Recurrent Neural Networks (RNNs)} %
RNNs and their variants, such as Long Short-Term Memory (LSTM) networks, are powerful in capturing temporal patterns and sequence dependencies in traffic flows. These models are particularly suited for time-series DDoS detection, where packet arrival times or flow durations evolve over time. However, in static feature selection tasks like identifying key features from tabular summaries of network flows. RNNs are less appropriate due to their sequential nature and higher computational cost. Additionally, RNNs provide limited feature-level interpretability without complex attention mechanisms.

\subsubsection{Transformer-Based Models} %
Transformers, known for their self-attention mechanism, have recently been explored for cybersecurity tasks, offering state-of-the-art performance in various sequence modelling problems. They can model long-range dependencies and automatically assign attention scores to inputs, which can be repurposed for feature importance estimation. However, their high computational demands, large data requirements, and complexity make them difficult to train on tabular datasets like CIC-DDoS2019 without significant architecture adjustments or pretraining. Furthermore, attention weights do not always correlate with actual feature importance \cite{goodfellow2020generative}.

Recent advancements in AI-based cybersecurity solutions leverage automated and intelligent feature selection techniques to enhance attack prediction capabilities. These AI-driven models, such as deep learning, ensemble methods, and intelligent agent systems, optimize feature selection dynamically, ensuring that only the most relevant attributes are used for real-time DDoS mitigation. Moreover, traditional filter, wrapper, embedded, and hybrid feature selection techniques, while effective in static environments, often struggle with the evolving nature of cyber threats. To address this, researchers have explored automated feature selection models, which integrate feature engineering, selection, and classification into a unified system. These intelligent systems reduce manual intervention and improve adaptability to new attack patterns. The following section explores how AI-based methods have been applied to DDoS detection, emphasizing automated selection, scalability, and real-time attack mitigation.

\subsection{Existing AI Based DDoS Detection Methods} %

\subsubsection{Automated and Intelligent Systems} %
Recent research has increasingly focused on developing automated systems that integrate feature extraction, selection, and classification into unified frameworks to enhance DDoS attack detection. These systems aim to reduce manual intervention, improve accuracy, and adapt to evolving attack patterns. \cite{bakar2023intelligent} proposes an intelligent agent system that automates feature extraction and selection, achieving a 99.7\% improvement in detection accuracy. \cite{ma2023ddos} introduces a framework (FAMS) that automates feature and model selection, aiming for high generalization capability and short prediction times.

\subsubsection{Performance Optimization and Scalability} %
With the increasing volume and complexity of network traffic data, optimizing DDoS detection systems for scalability and efficiency has become critical. Researchers have explored methods to handle large datasets without compromising detection accuracy. The study \cite{aamir2019ddos} demonstrates that feature reduction can significantly improve system efficiency without substantial performance degradation. By strategically reducing features, detection systems achieve faster processing speeds while retaining critical information. The strength of this approach lies in its simplicity and effectiveness, making it applicable to a wide range of scenarios. However, the process of identifying and removing redundant features requires domain expertise and careful analysis, which may not always be feasible in dynamic environments. Similarly, \cite{golduzian2023predict} addresses scalability challenges using advanced models like CNN and XGBoost. This approach effectively handles the CICDDoS2019 dataset, which contains over 50 million records, ensuring both accuracy and computational efficiency. The strength of these models lies in their ability to process large volumes of data with high precision, making them ideal for real-time detection. However, their reliance on extensive computational resources and training data may limit their applicability in environments with limited infrastructure. These efforts underscore the importance of balancing feature richness with system performance in large-scale network environments, while also highlighting the trade-offs between computational efficiency and resource requirements.

\subsubsection{Comparative Studies and Benchmarking} %
Comparative studies provide valuable insights into the performance of different feature selection methods and classification algorithms, establishing benchmarks for future research. Filter-based, wrapper-based, and embedded-based approaches are evaluated in the research \cite{kurniawan2022comparison}, which also highlights the accuracy, efficiency, and interpretability trade-offs associated with each. Filter-based techniques, such mutual information and correlation, are appropriate for initial feature screening since they are simple to use and computationally efficient. They could, however, fail to notice intricate relationships between characteristics, producing less-than-ideal outcomes. Wrapper-based methods, such as recursive feature elimination, offer higher accuracy by evaluating feature subsets based on model performance. Their weakness lies in their computational intensity, which can be prohibitive for large datasets. Embedded methods, such as Lasso regularization achieves equilibrium by including feature selection into the model training procedure. Despite their effectiveness and efficiency, the intricacy of the underlying algorithms may restrict their interpretability. In 2022, different research \cite{wang2022curse} looks at the connection between prediction accuracy and feature count. It demonstrates that while additional features can improve accuracy, they also introduce computational overhead, emphasizing the need for a balanced approach. For instance, decision trees and random forests perform well with smaller feature sets but struggle with high-dimensional data due to increased complexity. In contrast, neural networks and ensemble methods like XGBoost excel with larger feature sets but require significant computational resources and training time. To better understand the impact of different feature selection methods, a comparative analysis is conducted, evaluating their strengths, weaknesses, and suitability for DDoS detection. A systematic comparison of several filter, wrapper, embedding, and hybrid feature selection techniques is given in the following table, along with information on each method's benefits and drawbacks. This analysis helps in identifying the most efficient techniques for reducing feature redundancy while maintaining high detection accuracy. The next section presents a detailed comparison of existing feature selection methods, showcasing their impact on DDoS detection performance.

\begin{table}[H] %
 \caption{Strengths and Limitations of Feature Selection Methods in DDoS Detection}
 \centering
 \begin{tabular}{p{0.2\linewidth} p{0.2\linewidth} p{0.25\linewidth} p{0.25\linewidth}}
  \toprule
  Research Paper & FS Method & Strengths & Weaknesses \\
  \midrule
  Comparison of Feature Selection Methods for DDoS Attacks on Software Defined Networks using Filter- Based, Wrapper- Based and Embedded- Based\cite{kurniawan2022comparison} & Filter- based, Wrapper- based, Embedded- based & Identifying key features like 'pktcount' can lead to lightweight and early DDoS detection in SDNs. & The effectiveness depends on the quality and representativeness of the training data. Insufficient or biased data can lead to poor detection performance. \\
  \\
  Analysis of Machine Learning Classifiers for Early Detection of DDoS Attacks on IoT Devices\cite{gaur2021analysis} & Chi- Square, Extra Trees Classifier, ANOVA & Effective for categorical data, Captures complex interactions between features, identifies features that impact the target variable & Assumes independence between features, Prone to overfitting, sensitive to outliers \\
  \\
  Feature- Selection- Based DDoS Attack Detection Using AI Algorithms
  \cite{raza2024feature} & NGBoost, CNN & More Accuracy, less computational complexity & Potential overfitting, risk of omitting important features \\
    \\
  Curse of Feature Selection: a Comparison Experiment of DDoS Detection Using Classification Techniques
  \cite{wang2022curse} & Mutual Information & Simple and Efficient & Ignores feature interactions, sensitive to data distribution \\
  \\

  \bottomrule
 \end{tabular}
 \label{tab:fs_comparison}
\end{table}

\begin{table}[H] %
 \centering
 \begin{tabular}{p{0.2\linewidth} p{0.2\linewidth} p{0.25\linewidth} p{0.25\linewidth}}
  \toprule
  Research Paper & FS Method & Strengths & Weaknesses \\
  \midrule

   Feature Selection for DDoS Detection Using Classification Machine Learning Techniques
  \cite{maslan2020feature} & Linear Regression with the Forward Method & Easy to implement. Less risk of overfitting & Does not find optimal set of features, can include irrelevant features due to overfitting \\
  \\
An Intelligent Agent-Based Detection System for DDoS Attacks Using Automatic Feature Extraction and Selection
  \cite{bakar2023intelligent} & Intelligent Agent System, Sequential Feature Selection & Dynamic Adaptation, Enhanced Detection Performance (99.7\% improvement) & Computational complexity, dependence on data quality \\
  \\
  FSMDAD: Feature Selection Method for DDoS Attack Detection
  \cite{gaur2022fsmdad} & Chi- Square, Extra Trees Classifier, ANOVA, Mutual Information & Comprehensive evaluation, improved detection performance & Computational complexity, potential redundancy \\
  \\
  Defensive Mechanism Against DDoS Attack Based on Feature Selection and Multi-Classifier Algorithms
  \cite{mishra2022defensive} & Combined feature selection with multi- classifier systems (e.g., ensemble methods) & Enhanced robustness through classifier diversity. & Increased system complexity and computational costs. \\
  \\
  DDoS Attack Detection with Feature Engineering and Machine Learning: The Framework and Performance Evaluation
  \cite{aamir2019ddos} & Improved Binary Grey Wolf Optimization (wrapper- based method) & Achieves optimal feature subset selection, improving detection accuracy and reducing computational overhead. & Computationally intensive due to iterative search for optimal features, which may not scale well for large datasets. \\
  \\
  A DDoS Attack Detection Method Based on Natural Selection of Characteristics and Model Selection
  \cite{ma2023ddos} & Hybrid approach combining filter, wrapper, and embedded methods & Balances computational efficiency (filter methods) with model- specific optimization (wrapper/embedded methods) & Increased complexity in integrating multiple feature selection techniques \\
  \\
  Predict and Prevent DDoS Attacks Using Machine Learning and Statistical Algorithms
  \cite{golduzian2023predict} & Synthetic Minority Oversampling Technique (SMOTE) combined with statistical filtering methods. & Effectively addresses data imbalance and selects relevant features for improved model performance. & Dependence on SMOTE may lead to overfitting if not carefully tuned, especially in highly imbalanced datasets. \\
  \bottomrule
 \end{tabular}
 \label{tab:fs_comparison_2}
\end{table}

\section{Methodology} %
This section details the novel Generative Adversarial Network-based Feature Selection (GANFS) method developed for enhancing DDoS attack detection, the dataset utilized for evaluation, and the approach for setting key hyperparameters.

\subsection{Why GANs Were the Right Choice} %
The decision to use Generative Adversarial Networks (GANs) for feature selection was guided by both theoretical and practical considerations. Unlike other deep learning models that focus on classification, reconstruction, or sequence modelling, GANs are trained through an adversarial process in which the Generator learns to produce synthetic data that mimics the real data distribution, while the Discriminator learns to differentiate between real and fake samples \cite{goodfellow2020generative}. In this context, the Discriminator acts as a natural evaluator of feature relevance, its ability to detect fake attack samples hinges on the most critical features used to identify real ones \cite{maslan2020feature}. By performing a perturbation-based sensitivity analysis on the Discriminator's confidence scores, GANFS offers a principled, unsupervised method for ranking feature importance. Moreover, GANs are uniquely capable of modelling complex, non-linear relationships in high-dimensional data, making them ideal for capturing subtle patterns that might escape simpler models. Unlike CNNs or RNNs, GANFS does not require time-series or spatial feature encoding, and unlike Autoencoders, it focuses on discriminative power rather than reconstruction. Compared to Transformers, GANs offer a better balance of performance and computational efficiency for this problem setup \cite{wang2022curse}. In summary, although alternative deep learning architectures offer valuable capabilities for network security, GANs provide a direct, interpretable, and efficient mechanism for feature selection by leveraging the adversarial interplay between generative and discriminative learning. This makes GANFS particularly well-suited for high-dimensional DDoS detection scenarios where labeled data is limited and feature redundancy is high.

\subsection{GAN-Based Feature Selection (GANFS) Algorithm} %
The core challenge in DDoS detection using machine learning lies in the high dimensionality and redundancy often present in network traffic data. Traditional feature selection methods may struggle with complex feature interactions or require labelled data encompassing all traffic types. To address this, we propose a novel feature selection technique, GANFS, which leverages the discriminative power learned by a Generative Adversarial Network (GAN) trained specifically on attack patterns \cite{mishra2022defensive}.

\subsubsection{Rationale and Core Concept} %
The fundamental idea behind GANFS is to utilize the adversarial training dynamic of a GAN to identify features most critical for characterizing DDoS attacks. We train a GAN exclusively on samples representing various DDoS attacks (Label = 1). The GAN consists of two components:
\begin{itemize}
    \item \textbf{Generator (G)}: Takes random noise ($z$) as input and attempts to generate synthetic data samples that mimic the statistical properties of the real DDoS attack data.
    \item \textbf{Discriminator (D)}: Takes either a real DDoS attack sample ($x$) or a synthetic sample generated by G ($G(z)$) as input and tries to classify it as real or fake.
\end{itemize}
The two networks are trained adversarial. The Discriminator aims to maximize its accuracy in distinguishing real from fake samples, while the Generator aims to produce samples that are realistic enough to fool the Discriminator. This process is typically optimized using a value function like the one proposed by Goodfellow et al. \cite{goodfellow2020generative}:
\begin{equation} %
 \min_{G} \max_{D} V(D, G) = \mathbb{E}_{x \sim p_{data}(x)}[\log D(x)] + \mathbb{E}_{z \sim p_z(z)}[\log(1 - D(G(z)))]
\end{equation}
In the GANFS approach, we hypothesize that the features the trained Discriminator relies on most heavily to differentiate real attack samples from sophisticated fakes (generated by a well-trained Generator) are the features most salient and informative for describing the essential characteristics of the DDoS attacks themselves \cite{raza2024feature}.

\subsubsection{GAN Architecture} %
The specific GAN architecture employed in this study consists of Multi-Layer Perceptrons (MLPs) for both the Generator and the Discriminator:
\begin{itemize}
    \item \textbf{Generator}: Accepts a random noise vector (latent space) with dimensionality equal to the number of input features. It processes this noise through two dense hidden layers (64 and 128 neurons, respectively) using the Rectified Linear Unit (ReLU) activation function. The output layer consists of neurons equal to the number of features (81 in this case), using a Sigmoid activation function to ensure the generated feature values are scaled between 0 and 1, matching the normalized real data.
    \item \textbf{Discriminator}: Accepts an input vector representing a data sample (either real or fake) with 81 features. It passes this input through two dense hidden layers (128 and 64 neurons, respectively) with ReLU activation. The final output layer consists of a single neuron with a Sigmoid activation function, producing a probability score between 0 (interpreted as fake) and 1 (interpreted as real) \cite{mishra2022defensive}.
\end{itemize}

\subsubsection{GAN Training} %
The GAN was trained using only the pre-processed DDoS attack samples from the CIC-DDoS2019 dataset (details in Section 3.2). The training process involved:
\begin{itemize}
    \item \textbf{Data}: Using only attack samples (Label=1), normalized to the [0, 1] range.
    \item \textbf{Adversarial Loop}:
    \begin{itemize}
        \item \textit{Discriminator Training}: The Discriminator was trained on batches containing a mix of real attack samples (labelled close to 1, e.g., 0.9 for label smoothing) and fake samples generated by the Generator (labelled close to 0, e.g., 0.1).
        \item \textit{Generator Training}: The Generator was trained based on the Discriminator's feedback. Noise was fed to the Generator, the output was passed to the Discriminator, and the Generator's weights were updated to minimize the difference between the Discriminator's output and the target label 1 (i.e., aiming to fool the Discriminator).
    \end{itemize}
    \item \textbf{Optimization}: Both models were compiled using the Adam optimizer and Binary Cross entropy loss function.
    \item \textbf{Training Parameters}: Training was conducted for 500 epochs with a batch size of 4096.
\end{itemize}

\subsubsection{Feature Importance via Sensitivity Analysis (Perturbation Strategy)} %
Once the GAN is trained, the Generator is discarded, and the trained Discriminator is used for feature importance ranking. The importance of each feature is assessed via a sensitivity analysis based on systematic perturbation \cite{aamir2019ddos}:
\begin{itemize}
    \item \textbf{Baseline Confidence ($D(x)$)}: The baseline prediction confidence of the Discriminator is computed for the real attack samples.
    \item \textbf{Base Delta Calculation ($\Delta_{base}$)}: For each feature $i$, a characteristic step size $\Delta_{base_i}$ is calculated. This represents a meaningful minimum change for that feature, derived from the data distribution (e.g., calculated as the mean of non-zero differences between consecutive sorted values of that feature across the attack samples).
    \item \textbf{Perturbation Levels}: A set of perturbation factors $F = \{0.5, 1.0, 2.0, 5.0, 10.0\}$.
    \item \textbf{Bidirectional Perturbation}: For each feature $i$, each sample $x$, and each factor $f$ in $F$:
    \begin{itemize}
        \item Calculate perturbation magnitude: $\Delta_f = f \cdot \Delta_{base_i}$.
        \item Create two perturbed samples with each value of $\Delta_f$:
          \begin{itemize}
              \item Positive Perturbation: $x_{i}^{+} = x_i + \Delta_f$.
              \item Negative Perturbation: $x_{i}^{-} = x_i - \Delta_f$.
          \end{itemize}
        \item Apply clipping to ensure perturbed feature values remain within the valid normalized range [0, 1].
    \end{itemize}
    \item \textbf{Measure Confidence Change}: Calculate the absolute change in the Discriminator's output confidence for each perturbation:
    \begin{itemize}
        \item $\Delta_{\text{conf}}^{+} = |D(x) - D(x^{+})|$.
        \item $\Delta_{\text{conf}}^{-} = |D(x) - D(x^{-})|$.
    \end{itemize}
    \item \textbf{Aggregate Sensitivity}: For each feature $i$, the final sensitivity score is calculated by averaging the confidence changes over all samples, both perturbation directions (+/-), and all perturbation factors $f$ in $F$:
    \begin{equation} %
    \text{Sensitivity}_i = \frac{1}{N \cdot K \cdot 2} \sum_{\text{samples}} \sum_{f \in F} (\Delta_{\text{conf}}^{+} + \Delta_{\text{conf}}^{-})
    \end{equation}
     where N is the number of samples and K is the number of perturbation factors.
\end{itemize}
A higher sensitivity score indicates that perturbations in that feature cause larger fluctuations in the Discriminator's confidence, implying the feature is more critical for distinguishing real attacks from generated fakes and, thus, more important for characterizing the attack.

\subsubsection{Output} %
The GANFS method outputs a ranked list of all input features, ordered from most important (highest sensitivity score) to least important. This ranked list can then be used to select the top-k features for training downstream DDoS detection classifiers.

\subsection{Dataset Description} %
The dataset used for training the GANFS model and benchmarking the feature selection methods is the CIC-DDoS2019 dataset \cite{ddos2019dataset}. This dataset is widely recognized for evaluating DDoS detection mechanisms and contains a diverse range of modern reflection-based DDoS attack traffic alongside benign traffic.

\subsubsection{Composition and Size} %
\begin{itemize}
    \item \textbf{Source Files}: We utilized specific attack files from the dataset, including DrDoS\_LDAP.csv, DrDoS\_MSSQL.csv, DrDoS\_NetBIOS.csv, DrDoS\_NTP.csv, DrDoS\_SNMP.csv, DrDoS\_SSDP.csv, DrDoS\_UDP.csv, and DrDoS\_DNS.csv. These were merged with benign traffic samples present within them.
    \item \textbf{Initial Size}: The combined raw dataset comprised approximately 28 million records.
    \item \textbf{Label Distribution}: The dataset is highly imbalanced, containing roughly 1 million BENIGN samples and 27 million DDoS attack samples across the various types.
    \item \textbf{Sampling (for manageability)}: To manage computational resources during initial loading and processing, a subset of the attack data was potentially sampled (as indicated in one version of the loading code, e.g., 500,000 samples per attack file type were retained along with all benign samples, leading to approx. 4 million records used in the final ddos\_df). The final dataset used for splitting into train/test contained approximately 4,027,519 records.
\end{itemize}

\subsubsection{Features} %
\begin{itemize}
    \item \textbf{Initial Features}: The raw dataset contains over 80 features extracted from network flows using the CICFlowMeter tool.
    \item \textbf{Preprocessing and Feature Count}: After preprocessing (detailed below), 81 relevant features remained for analysis. These features encompass various aspects of network flows, including: Basic flow identifiers, Packet counts and lengths, Flow timing characteristics, TCP Flags counts, Payload-related features, Window size information, Flow rate features, Subflow averages, Active/Idle time statistics.
\end{itemize}

\subsubsection{Preprocessing Steps} %
Several preprocessing steps were applied to prepare the data:
\begin{itemize}
    \item \textbf{Merging}: Data from the selected CSV files were concatenated into a single Pandas DataFrame.
    \item \textbf{Irrelevant Feature Removal}: Columns deemed non-informative or problematic for modeling were dropped. These included Timestamp, Source IP, Destination IP, Flow ID, SimillarHTTP, and any unnamed index columns (Unnamed: 0). Column names were stripped of leading/trailing whitespace.
    \item \textbf{Handling Invalid Values}: Infinite values (Infinity, np.inf, -Infinity) and Not-a-Number (NaN) values, particularly arising from division by zero in rate calculations (e.g., Flow Packets/s, Flow Bytes/s), were replaced with 0.
    \item \textbf{Label Encoding}: The categorical 'Label' column was converted into a binary numerical format: BENIGN was mapped to 0, and all specific DrDoS attack types were mapped to 1.
    \item \textbf{Data Filtering (for GANFS)}: Crucially, for training the GANFS model itself, the dataset was filtered to retain only attack samples (where Label = 1). The benchmarking classifiers were trained and tested on the full dataset containing both benign and attack samples.
    \item \textbf{Normalization}: All 81 numerical feature columns were scaled to the range [0, 1] using sklearn.preprocessing.MinMaxScaler. This ensures all features contribute uniformly during GAN training and prevents features with larger ranges from dominating the learning process.
\end{itemize}

\subsection{Hyperparameter Setting Methodology} %
Setting appropriate hyperparameters is crucial for the performance and stability of the GANFS model and the benchmark classifiers \cite{ma2023ddos}. The following methodology was employed:

\subsubsection{GANFS Model Hyperparameters} %
The hyperparameters for the GAN (Generator and Discriminator) were selected based on a combination of common practices in GAN literature for tabular data, preliminary experiments to ensure training stability, and resource constraints.
\begin{itemize}
    \item \textbf{Network Architecture}: The number of layers (2 hidden layers) and neurons (64/128 for G, 128/64 for D) were chosen as a standard MLP configuration. ReLU activation was used for hidden layers, while Sigmoid was used for output layers.
    \item \textbf{Optimizer}: The Adam optimizer was selected.
    \item \textbf{Learning Rate}: A learning rate of 0.001 was used. No extensive learning rate search was performed.
    \item \textbf{Loss Function}: Binary Cross entropy was used.
    \item \textbf{Epochs}: The model was trained for 500 epochs, determined by monitoring loss curves.
    \item \textbf{Batch Size}: A batch size of 4096 was used for gradient stability and resource management.
\end{itemize}

\subsubsection{Sensitivity Analysis Parameters} %
\begin{itemize}
    \item \textbf{Perturbation Factors}: The factors $\{0.5, 1.0, 2.0, 5.0, 10.0\}$ were chosen empirically to cover a range of magnitudes relative to feature granularity.
    \item \textbf{Base Delta Calculation}: Using the mean of non-zero differences between sorted consecutive values provides a data-driven estimate of a meaningful small change.
\end{itemize}

\subsubsection{Benchmark Classifier Hyperparameters} %
For the benchmark classifiers (Logistic Regression, Random Forest) used to evaluate the selected feature subsets:
\begin{itemize}
    \item \textbf{Logistic Regression}: The max\_iter parameter was set to 1000 to ensure convergence. Other hyperparameters were kept at their default values.
    \item \textbf{Random Forest}: Default scikit-learn hyperparameters were used.
\end{itemize}
It is important to note that extensive hyperparameter optimisation (e.g., via grid search or randomised search) for the benchmark classifiers was not performed in this phase of the research. The focus was on comparing the effectiveness of different feature sets derived from various selection methods using reasonably configured standard classifiers. Fine-tuning classifier hyperparameters is considered future work.

\section{Experimental setup} %
The experiments were conducted on a high-performance computing node equipped with the following specifications:
\begin{itemize}
    \item CPU: Intel(R) Xeon(R) Gold 6326 CPU @ 2.90GHz 
    \item GPU: NVIDIA H100 with 10 GB of dedicated VRAM 
    \item System RAM: 100 GB 
\end{itemize}

\section{Results} %

\subsection{Features obtained from GANFS} %
Top 20 features obtained from our new GAN Feature Selection technique are URG Flag Count, Protocol, Inbound, Bwd Packet Length Max, Down/Up Ratio, Idle Std, RST Flag Count, Fwd PSH Flags, Bwd IAT Min, Active Mean, SYN Flag Count, Total Backward Packets, CWE Flag Count, act\_data\_pkt\_fwd, Bwd Packet Length Min, Init\_Win\_bytes\_backward, Init\_Win\_bytes\_forward, ACK Flag Count, Bwd Packet Length Mean, Active Std.
\\
\textit{Note: Full ranking of all 81 features along with their sensitivity score is included in the appendix.}

\subsection{Interpretation from the results} %
\begin{itemize}
    \item URG Flag Count was identified as the most sensitive feature, suggesting it contributes the most to the model’s decision-making.
    \item Features like Protocol, Inbound, and Bwd Packet Length Max also shows strong influence.
    \item Several features had zero sensitivity, indicating they can potentially be excluded to optimize performance.
\end{itemize}

\subsection{Evaluation Metrics} %
In this study, multiple performance metrics were evaluated to assess the effectiveness of the Generative Adversarial Network-based Feature Selection (GANFS) method for detecting DDoS attacks. The key metrics considered include:
\begin{itemize}
    \item \textbf{Accuracy}: Measures the overall performance by computing the ratio of correct predictions to the total number of predictions. It shows how well the model generalizes across the entire dataset.
    $$ \text{Accuracy} = \frac{TP+TN}{TP+TN+FP+FN} $$ 
    \item \textbf{Precision}: Focuses on the correctly predicted DDoS attacks relative to all predicted DDoS attacks. Crucial for reducing false positives.
    $$ \text{Precision} = \frac{TP}{TP+FP} $$ 
    \item \textbf{Recall (Sensitivity)}: Measures the model's ability to correctly identify actual DDoS attacks. Important for minimizing false negatives.
    $$ \text{Recall} = \frac{TP}{TP+FN} $$ 
    \item \textbf{F1-Score}: The harmonic mean of precision and recall, providing a balance, especially in imbalanced classes.
    $$ F1 = 2 \times \frac{\text{Precision} \times \text{Recall}}{\text{Precision} + \text{Recall}} $$ 
    \item \textbf{AUC-ROC Curve}: Illustrates the model's ability to distinguish between classes at various thresholds by plotting True Positive Rate vs. False Positive Rate. The Area Under the Curve (AUC) quantifies this ability.
    \item \textbf{Computational Efficiency}: Measured by training time and resource consumption, essential for real-time detection.
\end{itemize}

\subsection{Testing Methodology} %
The testing methodology adopted for this study involved the following:
\begin{itemize}
    \item \textbf{Training and Testing Split}: The dataset was split into training and testing subsets to evaluate generalization to unseen data.
    \item \textbf{Sensitivity Analysis}: GANFS was evaluated based on the sensitivity analysis of selected features by perturbing each feature to quantify its influence. This identified significant features and improved interpretability.
    \item \textbf{Baseline Comparisons}: GANFS was compared to traditional filter, wrapper, and embedded methods to demonstrate its advantages.
\end{itemize}
The testing methodology is appropriate as it assesses classification accuracy, computational efficiency, and feature importance through sensitivity analysis, making it suitable for evaluating real-time DDoS detection capabilities.

\subsection{Baseline Methods} %
To evaluate the performance of the proposed GANFS method, it was compared against several baseline and competitor methods commonly used in DDoS detection. These methods include:
\begin{itemize}
    \item \textbf{Filter-based Methods}:
    \begin{itemize}
        \item \textit{Mutual Information (MI)}: Measures feature-target dependency. Computationally efficient but ignores feature interactions.
        \item \textit{Chi-Square Test}: Assesses categorical feature-target relationship. Useful but also ignores interactions.
    \end{itemize}
    \item \textbf{Wrapper-based Methods}:
    \begin{itemize}
        \item \textit{Recursive Feature Elimination (RFE)}: Iteratively removes features based on model performance. Accurate but computationally expensive due to repeated model training.
    \end{itemize}
    \item \textbf{Embedded Methods}:
    \begin{itemize}
        \item \textit{Random Forest Feature Importance}: Calculates importance based on feature influence in decision trees. Efficient and handles high dimensions.
    \end{itemize}
    \item \textbf{Competitor Classifiers}:
    \begin{itemize}
        \item \textit{Logistic Regression}: Simple, interpretable baseline classifier.
        \item \textit{Random Forest Classifier}: More complex, robust ensemble classifier.
    \end{itemize}
\end{itemize}

\subsubsection{Rationale for Selection} %
These baseline methods were selected to represent a variety of feature selection techniques and classification models, from simple to complex. The goal was to compare GANFS against well-established methods to highlight its strengths, particularly in computational efficiency and feature importance identification. This comparison helps gauge the improvement GANFS offers over traditional approaches.

\subsection{Description of Results} %
The results demonstrated the effectiveness of GANFS in detecting DDoS attacks, evaluated using accuracy, precision, recall, F1-score, and computational efficiency.

\subsubsection{Accuracy} %
\begin{figure}[h!]
  \centering
  \includegraphics[width=1\linewidth, height=0.2\linewidth]{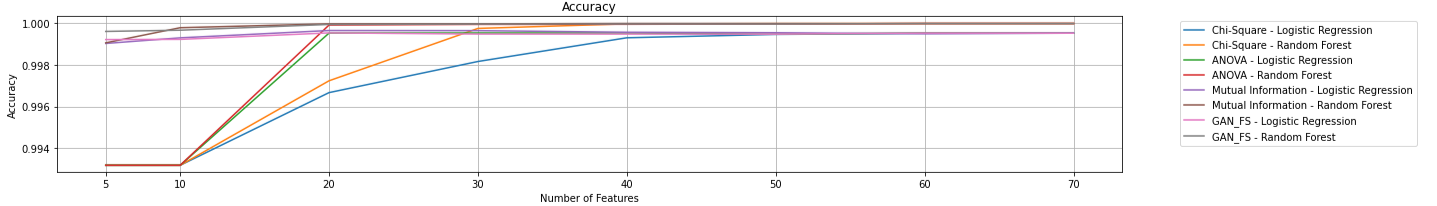}
  \caption{Accuracy achieved by different Feature Selection techniques for two models – Logistic Regression and Random Forest.}
  \label{fig:accuracy}
\end{figure}
GANFS performed comparably to baseline methods, achieving peak accuracies of 99.9995\% with Random Forest and 99.954\% with Logistic Regression. These were higher compared to Chi-Square (99.9995\%), ANOVA (99.9996\%), and Mutual Information (99.9998\%). Performance indicates GANFS's robustness.

\subsubsection{Precision} %
\begin{figure}[h!]
  \centering
  \includegraphics[width=1\linewidth, height=0.2\linewidth]{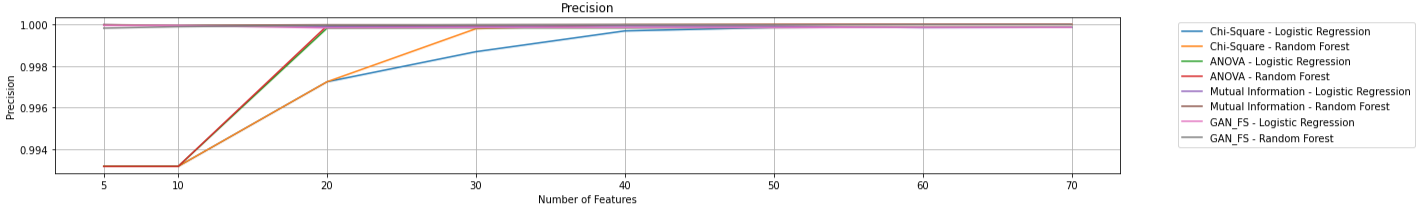}
  \caption{Precision score achieved by different Feature Selection techniques for two models – Logistic Regression and Random Forest.}
  \label{fig:precision}
\end{figure}
GANFS showed similar precision, achieving 99.9999\% with Random Forest and 99.9952\% with Logistic Regression. This surpassed other methods like Chi-Square (99.9999\%), ANOVA (99.9999\%), and Mutual Information (99.9999\%). Higher precision reduces false positives.

\subsubsection{Recall} %

\begin{figure}[H]
  \centering
  \includegraphics[width=1\linewidth, height=0.2\linewidth]{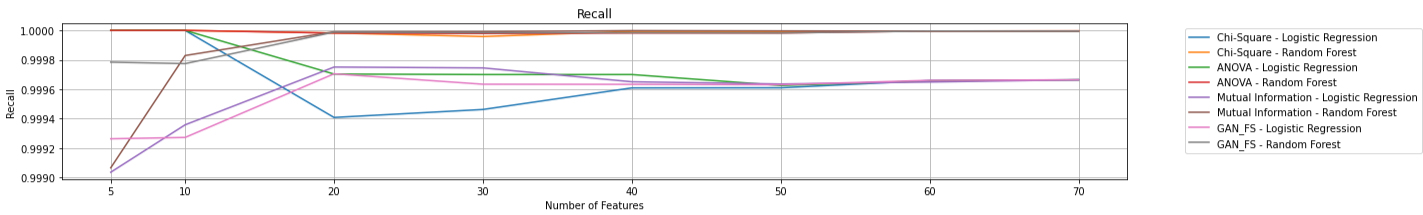}
  \caption{Recall of different Feature Selection techniques for two models – Logistic Regression and Random Forest.}
  \label{fig:recall}
\end{figure}

GANFS showed similar recall, with peak rates of 99.9996\% (Random Forest) and 99.9664\% (Logistic Regression). Comparable to Chi-Square (99.9997\%), ANOVA (99.9997\%), and Mutual Information (99.9999\%). High recall means effective detection of actual attacks.
\\
\subsubsection{F1 Score} %
\begin{figure}[h!]
  \centering
  \includegraphics[width=1\linewidth, height=0.2\linewidth]{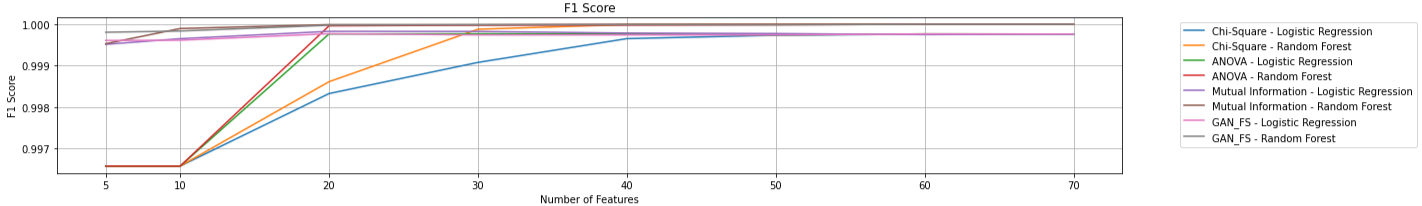}
  \caption{F1 Score achieved by different Feature Selection techniques for two models – Logistic Regression and Random Forest.}
  \label{fig:f1_score}
\end{figure}
GANFS achieved higher F1 scores (99.9997\% with Random Forest, 99.9769\% with Logistic Regression) compared to Chi-Square (99.9998\%), ANOVA (99.9998\%), and Mutual Information (99.9999\%). This shows a better balance between precision and recall, making GANFS more reliable.

\subsubsection{AUC-ROC Curve} %
\begin{figure}[h!]
  \centering
  \includegraphics[width=1\linewidth, height=0.2\linewidth]{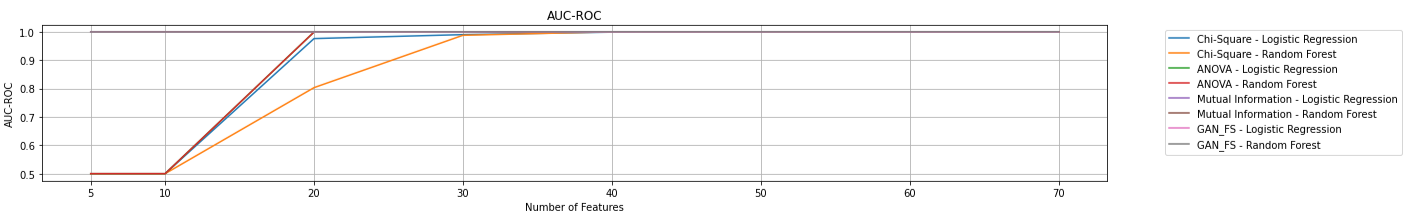}
  \caption{AUC-ROC Curve of different Feature Selection techniques for two models – Logistic Regression and Random Forest.}
  \label{fig:auc_roc}
\end{figure}
GANFS achieved significantly higher AUC-ROC scores (up to 0.99999999 with Random Forest, 0.99976 with Logistic Regression), surpassing baselines like Chi-Square (0.99976), ANOVA (0.99976), and Mutual Information (0.99999). This highlights superior discriminative ability.

\subsubsection{Computational Efficiency} %
\begin{figure}[h!]
  \centering
  \includegraphics[width=0.7\linewidth]{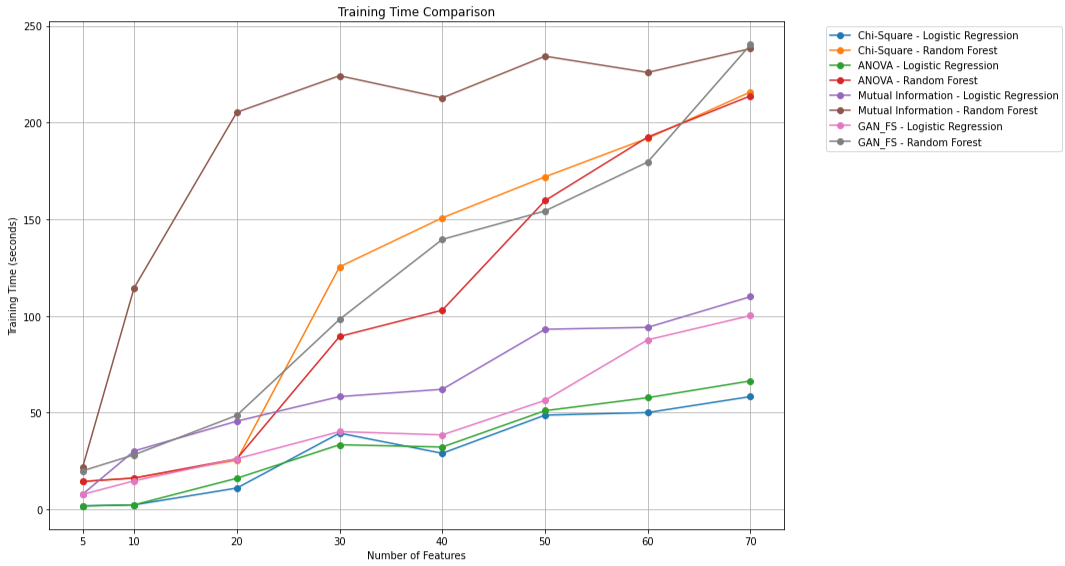}
  \caption{Computational Efficiency of different Feature Selection techniques for two models – Logistic Regression and Random Forest.}
  \label{fig:comp_eff}
\end{figure}
GANFS proved most efficient in training time compared to traditional methods. Results highlight GANFS scalability for large applications, crucial for real-time detection. Importance of balancing efficiency and accuracy is underscored.

\subsection{Performance comparison for Logistic Regression} %
    \begin{figure}[H]
      \centering
      \includegraphics[width=1\linewidth]{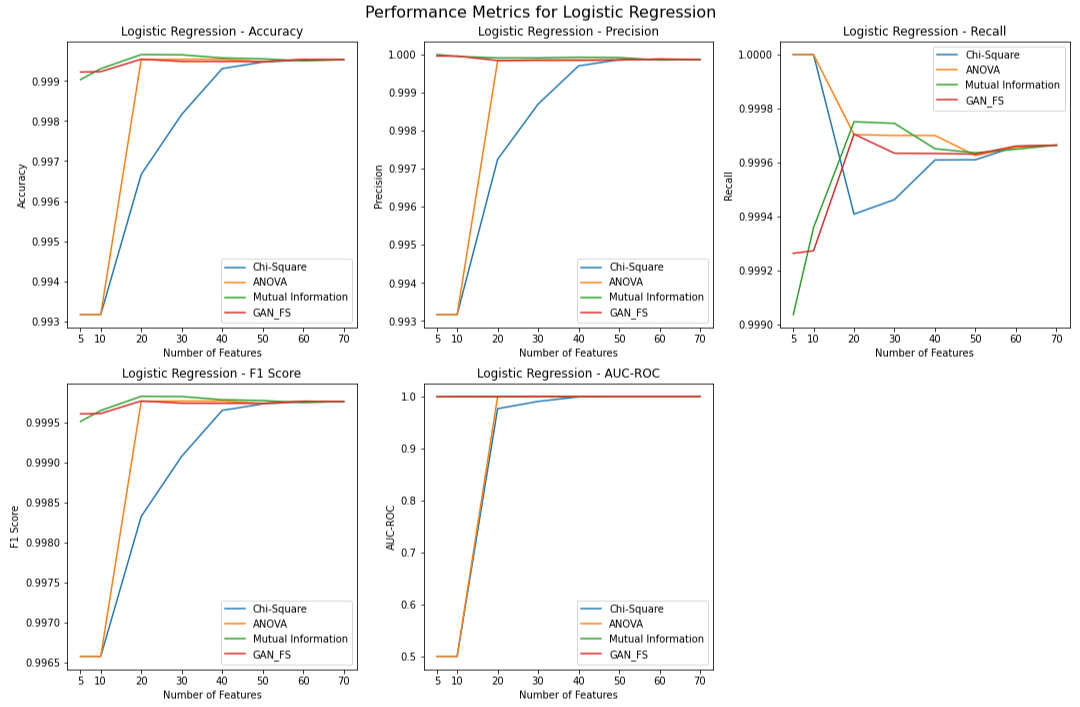}
      \caption{Performance Metrics by different Feature Selection techniques for Logistic Regression.}
      \label{fig:lr}
    \end{figure}

Accuracy: GANFS shows quick initial improvement, outperforming others early. ANOVA improves rapidly but doesn't surpass GANFS. Chi-Square and MI follow similar paths, slightly lower than GANFS. Conclusion: GANFS achieves optimal performance rapidly.

Precision: GANFS rapidly achieves peak performance. ANOVA catches up quickly. Chi-Square and MI improve steadily but don't match GANFS. Conclusion: GANFS outperforms, ANOVA trails closely.

Recall: Shows fluctuating, unstable pattern. GANFS and ANOVA peak sharply then fluctuate. MI starts lower, peaks mid-way, then stabilizes. Chi-Square decreases then stabilizes lower. Conclusion: Challenging data/scenario, GANFS strong initially but unstable in recall.

F1 Score: GANFS quickly reaches and maintains peak performance. ANOVA shows sharp improvement but slightly below GANFS. MI and Chi-Square track closely, MI slightly better. Conclusion: GANFS shows consistent superiority.

AUC-ROC: GANFS achieves max performance quickly, surpassing others. ANOVA rises sharply, slightly below GANFS. Chi-Square follows ANOVA closely. MI matches Chi-Square, lagging GANFS/ANOVA. Conclusion: GANFS maintains highest, quickest-achieved performance.

\begin{figure}[H]
  \centering
  \includegraphics[width=1\linewidth]{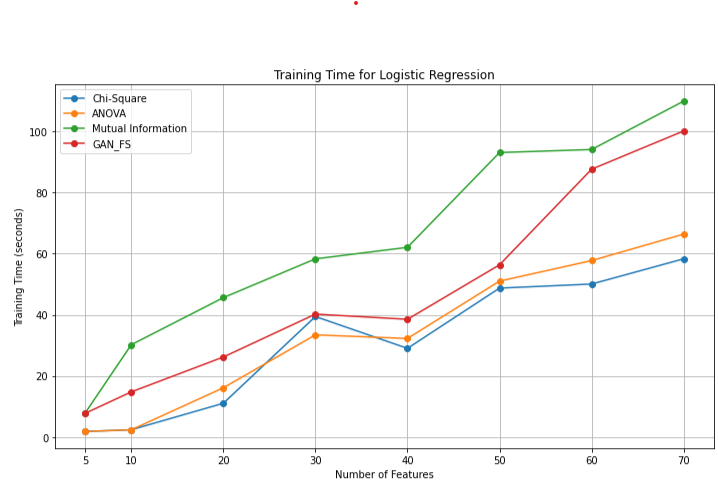}
  \caption{Computational Efficiency of different Feature Selection techniques for Logistic Regression.}
  \label{fig:lr_comp_eff}
\end{figure}
Computational Efficiency (Figure 8): Training time increases with features for all methods. ANOVA or Chi-Square preferable for rapid training. GANFS or MI beneficial for feature quality (higher cost), GANFS faster than MI at larger feature counts. Choice depends on balancing resources and objectives.

\subsection{Performance Metrics for Random Forest} %
\begin{figure}[H]
  \centering
  \includegraphics[width=1\linewidth]{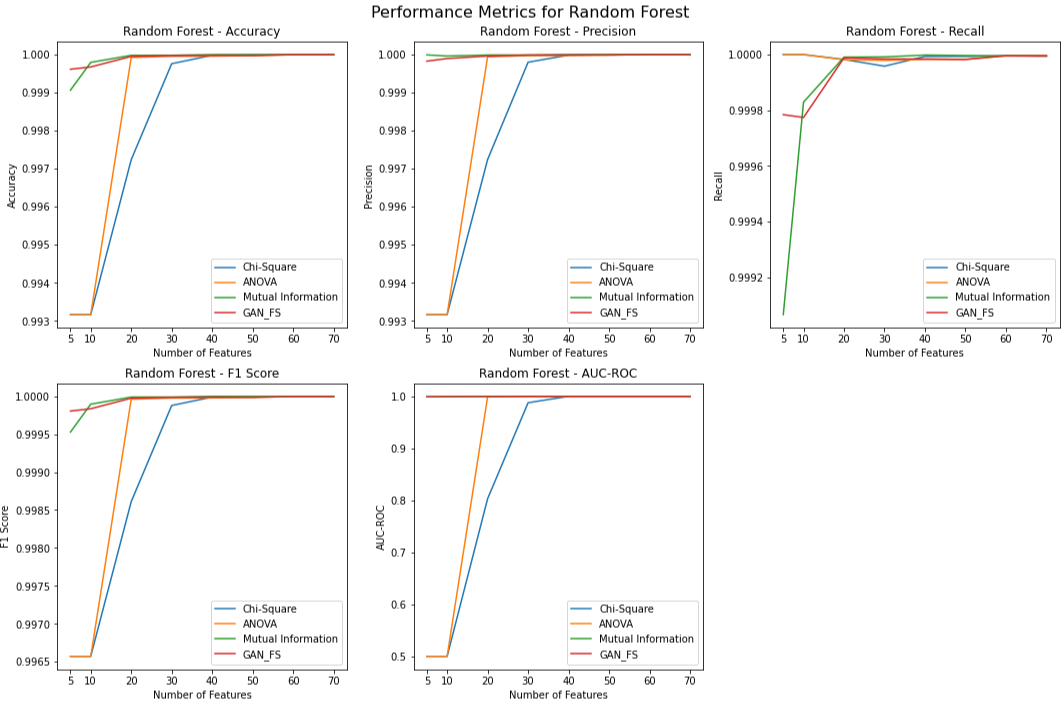}
  \caption{Accuracy achieved by different Feature Selection techniques for Random Forest. (Placeholder)}
  \label{fig:rf_accuracy}
\end{figure}
Accuracy: GANFS achieves highest performance immediately and maintains stability. MI closely follows, slightly lower. ANOVA improves rapidly but remains below GANFS/MI. Chi-Square improves gradually, similar to ANOVA but slower. Conclusion: GANFS superior rapid performance and MI is also strong.

Precision: GANFS achieves maximum performance immediately. ANOVA improves sharply, almost reaching GANFS later. Chi-Square improves gradually. MI consistently strong, tracking GANFS. Conclusion: GANFS and MI lead, GANFS slightly superior.

Recall: Slightly different trend. ANOVA starts high and maintains top performance, slightly outperforming GANFS. GANFS stabilizes high but slightly below ANOVA. MI and Chi-Square reach high performance quickly, tracking GANFS. Conclusion: ANOVA superior here, indicating dataset-dependent advantages.

F1 Score: GANFS clearly dominates, reaching peak immediately. MI closely follows. ANOVA improves sharply, matching GANFS eventually. Chi-Square improves slower. Conclusion: GANFS and MI highly effective, followed by ANOVA.

AUC-ROC: GANFS exhibits immediate and consistent peak performance. ANOVA catches up rapidly. Chi-Square lags slightly. MI likely follows similar high performance. Conclusion: GANFS dominant, ANOVA quickly follows.

\begin{figure}[H]
  \centering
  \includegraphics[width=1\linewidth]{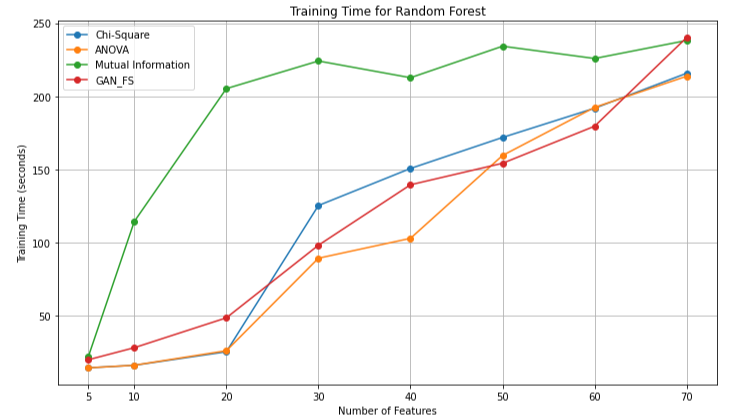}
  \caption{Computational Efficiency of different Feature Selection techniques for Random Forest. (Placeholder)}
  \label{fig:rf_comp_eff}
\end{figure}
Computational Efficiency (Figure 10): Training time increases with features for all. Chi-Square/ANOVA favourable for low feature counts. GANFS starts moderate, cost increases with dimensionality. MI consistently shows highest training times.

\subsection{Critical Discussion of Results} %
Our experiments showed GANFS achieved high accuracy, precision, and recall, especially with Random Forest classifiers. GANFS + RF consistently delivered F1 > 0.9999. Logistic Regression also performed exceptionally well. Results reflect the discriminator's ability to capture nuanced dependencies tied to malicious patterns. Compared to traditional methods, GANFS selected features providing more stable, higher performance across classifiers. This is attributed to model-driven feedback: the discriminator learns which variations influence confidence, leading to more task-relevant subsets. GANFS's data-driven adaptability, trained on diverse attack samples, likely contributed to good generalization.

Challenges encountered: GAN training is computationally demanding and time-intensive. Minor variations in importance rankings occurred across runs due to stochastic elements, though top features remained stable, suggesting robustness. Insightful finding: Some features highly ranked by traditional methods were down-ranked by GANFS, indicating statistical significance doesn't always equate to classification utility.

\subsection{Relevance of Results} %
Study aimed to improve DDoS detection by reducing features while maintaining performance. Results support this: classifiers using GANFS-selected features retained performance nearly identical to full sets. GANFS often outperformed traditional methods. Key insight: GANFS ranks based on actual contribution to model decisions (functional relevance), making selected features more robust and less noisy/redundant. Crucial for operational systems needing speed/efficiency. Approach proved effective across attack types/distributions, highlighting generalizability essential for evolving cyber threats.

Challenge identified: Potential use in dynamic/real-time environments. Applying GANFS live is problematic: selecting new features requires retraining the downstream model from scratch, as the classifier is coupled with the feature set. Retraining isn't always feasible in high-availability systems. Frequent updates increase overhead, downtime, and reduce responsiveness. The GAN itself isn't designed for incremental adaptation. Future work could explore hybrid approaches or continual learning.

\section{Conclusion} %
We proposed GANFS, a novel method using GAN discriminator sensitivity analysis to identify informative network features for DDoS detection. Evaluated on CIC-DDoS2019, GANFS outperformed traditional methods (Chi-Square, MI, ANOVA, RFE). With Logistic Regression, peak F1 was 0.9997 with Random Forest, performance was near-perfect (F1 > 0.9999, AUC-ROC 0.9999). GANFS proved computationally efficient, achieving top results with reduced features, lowering training/inference times. While Chi-Square/ANOVA were faster at low dimensions, GANFS balanced performance/cost better, especially at higher feature counts.

Key challenge: GANFS suits static/offline environments. Dynamic use requires retraining classifiers when features update, limiting real-time applicability without adaptation (e.g., incremental learning).

Overall, GANFS is a powerful, data-driven feature selection approach for DDoS detection, offering: (1) high accuracy with fewer features, (2) model-agnostic, unsupervised relevance estimation, and (3) improved scalability. Promising for next-gen IDS. Future work includes real-time integration and generalization to multiclass/multi-stage attacks.

\section{Future Work} %
While GANFS shows promise, several opportunities exist for expansion and enhancement.

\subsection{Expansion to Multiclass and Multi-Stage Attack Detection} %
This study focused on binary classification (DDoS vs. benign). CIC-DDoS2019 has multiple attack types. Future work: extend GANFS to multiclass to identify attack-specific features. Investigate temporal extensions (RNNs/GANs) for multi-stage attacks (reconnaissance then attack) to find time-dependent features \cite{kurniawan2022comparison, gaur2021analysis}.

\subsection{Evaluation Across Diverse Datasets} %
Validate GANFS generalizability beyond CIC-DDoS2019 using datasets like NSL-KDD, UNSW-NB15, or real enterprise logs. Apply to IoT datasets to test utility in resource-constrained environments. Assess consistency of selected features \cite{maslan2020feature, gaur2022fsmdad}.

\subsection{Integration with Online and Real-Time Detection} %
Current GANFS is offline/batch. Real-time systems need rapid inference/adaptation. Develop online/streaming GANFS updating importance over time. Incorporate continual learning for low-latency environments (edge/SDNs) \cite{subasri2022machine, bakar2023intelligent}.

\subsection{Comparative Study with Advanced Generative Models} %
Compare standard GAN with alternatives (VAEs, WGANs, Diffusion Models) for stability/representation. Study which generative mechanisms best facilitate feature discrimination. Explore transformer-based models with attention for embedded selection \cite{goodfellow2020generative, wang2022curse}.

\subsection{Ensemble Feature Ranking Strategies} %
Combine GANFS with traditional methods (MI, Chi-Square, RF) for ensemble ranking. Aggregate rankings (voting, fusion) for a robust, consensus-driven strategy. May mitigate individual weaknesses and improve stability \cite{mishra2022defensive, golduzian2023predict}.

\subsection{GANFS-Driven Hyperparameter Optimization} %
Explore interaction between GANFS-selected features and classifier hyperparameters. See if selected features yield better performance under optimized configurations (grid search, Bayesian optimization). Determine if FS amplifies or dampens tuning sensitivity \cite{raza2024feature, aamir2019ddos}.

\subsection{Interpretability and Explainability Enhancements} %
Enhance interpretability beyond sensitivity scores. Integrate SHAP/LIME for localized insights on feature influence. Use visualization to show perturbation effects on Discriminator confidence, improving transparency/trust \cite{ma2023ddos, goodfellow2020generative}.

\subsection{Broader Perspective and Recommendations} %
GANFS integrates deep generative modeling with intelligent FS for cybersecurity. Unsupervised, data-driven ranking suits real-world limits (labeled data scarcity). Sensitivity analysis framework is intuitive/generalizable beyond DDoS (fraud, medical, industrial anomaly).
Recommended strategic directions:
\begin{itemize}
    \item Adapt for multiclass/multi-domain settings \cite{maslan2020feature, bakar2023intelligent}.
    \item Optimize for real-time/online learning pipelines \cite{subasri2022machine, wang2022curse}.
    \item Integrate ensemble/interpretability methods for robustness/transparency \cite{mishra2022defensive, goodfellow2020generative}.
\end{itemize}
Pursuing these directions can evolve GANFS into a deployable component in modern security infrastructures.

\section*{Acknowledgments}
The author gratefully acknowledges the University of Ottawa for providing the computing hardware and facilities essential for this research. I would also like to express my sincere thanks to Dr.\ Paula Branco, Assistant Professor in the School of Electrical Engineering and Computer Science, for her invaluable guidance, insightful feedback, and continued support throughout this work.

\bibliographystyle{unsrt}

\begin{thebibliography}{10}

\bibitem{kurniawan2022comparison}
M.~Kurniawan, S.~Yazid, and Y.~G. Sucahyo.
\newblock Comparison of feature selection methods for ddos attacks on software defined networks using filter-based, wrapper-based and embedded-based.
\newblock {\em JOIV International Journal on Informatics Visualization}, 6(4):809, 2022. 

\bibitem{gaur2021analysis}
V.~Gaur and R.~Kumar.
\newblock Analysis of machine learning classifiers for early detection of ddos attacks on iot devices.
\newblock {\em Arabian Journal for Science and Engineering}, 47(2):1353--1374, 2021. 

\bibitem{raza2024feature}
M.~S. Raza, M.~N.~A. Sheikh, I.-S. Hwang, and M.~S. Ab-Rahman.
\newblock Feature-selection-based ddos attack detection using ai algorithms.
\newblock {\em Telecom}, 5(2):333--346, 2024.

\bibitem{wang2022curse}
W.~Wang, S.~M. Sadjadi, and N.~Rishe.
\newblock Curse of feature selection: a comparison experiment of ddos detection using classification techniques.
\newblock In {\em 2022 IEEE Intl Conf on Parallel \& Distributed Processing with Applications, Big Data \& Cloud Computing, Sustainable Computing \& Communications, Social Computing \& Networking (ISPA/BDCloud/SocialCom/SustainCom)}, pages 262--269, 2022. 

\bibitem{maslan2020feature}
A.~Maslan, K.~M.~B. Mohamad, and F.~B.~M. Foozy.
\newblock Feature selection for ddos detection using classification machine learning techniques.
\newblock {\em IAES International Journal of Artificial Intelligence}, 9(1):137, 2020.

\bibitem{subasri2022machine}
S.~I, E.~S.~G.~S. R., and R.~M.~P.
\newblock Machine learning based feature selection for ddos detection in named data networking.
\newblock In {\em 2022 4th International Conference on Advances in Computing, Communication Control and Networking (ICAC3N)}, pages 305--310, 2022. 

\bibitem{bakar2023intelligent}
R.~A. Bakar, X.~Huang, M.~S. Javed, S.~Hussain, and M.~F. Majeed.
\newblock An intelligent agent-based detection system for ddos attacks using automatic feature extraction and selection.
\newblock {\em Sensors}, 23(6):3333, 2023.

\bibitem{gaur2022fsmdad}
V.~Gaur and R.~Kumar.
\newblock Fsmdad: Feature selection method for ddos attack detection.
\newblock In {\em 2022 International Conference on Electronics and Renewable Systems (ICEARS)}, pages 939--944, 2022. 

\bibitem{mishra2022defensive}
A.~Mishra, N.~Gupta, and B.~B. Gupta.
\newblock Defensive mechanism against ddos attack based on feature selection and multi-classifier algorithms.
\newblock {\em Telecommunication Systems}, 82(2):229--244, 2022.

\bibitem{aamir2019ddos}
M.~Aamir and S.~M.~A. Zaidi.
\newblock Ddos attack detection with feature engineering and machine learning: The framework and performance evaluation.
\newblock {\em International Journal of Information Security}, 18(6):761--785, 2019.

\bibitem{ma2023ddos}
R.~Ma, X.~Chen, and R.~Zhai.
\newblock A ddos attack detection method based on natural selection of features and models.
\newblock {\em Electronics}, 12(4):1059, 2023.

\bibitem{golduzian2023predict}
A.~Golduzian.
\newblock Predict and prevent ddos attacks using machine learning and statistical algorithms.
\newblock {\em arXiv preprint arXiv:2308.15674}, 2023. 

\bibitem{goodfellow2020generative}
I.~Goodfellow, J.~Pouget-Abadie, M.~Mirza, B.~Xu, D.~Warde-Farley, S.~Ozair, A.~Courville, and Y.~Bengio.
\newblock Generative adversarial networks.
\newblock {\em Communications of the ACM}, 63(11):139--144, 2020.

\bibitem{ddos2019dataset}
University of New Brunswick.
\newblock Ddos 2019 dataset.
\newblock \url{https://www.unb.ca/cic/datasets/ddos-2019.html}. Accessed. 

\end{thebibliography}


\section*{Appendix} %

\begin{table}[h!] %
 \caption{Features ranked according to the sensitivity by GAN Feature Selection Algorithm}
 \centering
 \small 
 \begin{tabular}{rlr}
  \toprule
  S.No. & Feature & Sensitivity\_Score \\
  \midrule
  1 & URG Flag Count & 0.022578694 \\
  2 & Protocol & 0.016681384 \\
  3 & Inbound & 0.0154445 \\
  4 & Bwd Packet Length Max & 0.010720941 \\
  5 & Down/Up Ratio & 0.005439217 \\
  6 & Idle Std & 0.005243474 \\
  7 & RST Flag Count & 0.00336254 \\
  8 & Fwd PSH Flags & 0.003148519 \\
  9 & Bwd IAT Min & 0.002834816 \\
  10 & Active Mean & 0.002347851 \\
  11 & SYN Flag Count & 0.002086169 \\
  12 & Total Backward Packets & 0.00166628 \\
  13 & CWE Flag Count & 0.001646711 \\
  14 & act\_data\_pkt\_fwd & 0.001428789 \\
  15 & Bwd Packet Length Min & 0.001359528 \\
  16 & Init\_Win\_bytes\_backward & 0.001347067 \\
  17 & Init\_Win\_bytes\_forward & 0.001196553 \\
  18 & ACK Flag Count & 0.0011681 \\
  19 & Bwd Packet Length Mean & 0.001132366 \\
  20 & Active Std & 0.001072747 \\
  21 & Active Max & 0.001057202 \\
  22 & Subflow Bwd Bytes & 0.000883447 \\
  23 & Avg Bwd Segment Size & 0.000768293 \\
  24 & Active Min & 0.000595908 \\
  25 & Subflow Bwd Packets & 0.00056427 \\
  26 & Total Length of Bwd Packets & 0.000498126 \\
  27 & Bwd Packet Length Std & 0.000300974 \\
  28 & Fwd Packet Length Min & 0.000252315 \\
  29 & Bwd Header Length & 0.000226241 \\
  30 & Average Packet Size & 0.000194247 \\
  31 & Max Packet Length & 0.000183333 \\
  32 & Fwd IAT Min & 0.000146531 \\
  33 & Fwd Packet Length Std & 0.000139206 \\
  34 & Fwd Packet Length Mean & 0.000133515 \\
  35 & Bwd IAT Total & 0.000127808 \\
  36 & min\_seg\_size\_forward & 0.000111856 \\
  37 & Bwd IAT Std & 0.000110161 \\
  38 & Total Length of Fwd Packets & 9.98E-05 \\
  39 & Bwd IAT Max & 9.94E-05 \\
  40 & Subflow Fwd Bytes & 8.30E-05 \\
  41 & Fwd Packet Length Max & 8.06E-05 \\
  42 & Idle Max & 7.02E-05 \\
  43 & Subflow Fwd Packets & 6.64E-05 \\
  44 & Packet Length Variance & 6.03E-05 \\
  45 & Idle Mean & 5.62E-05 \\
  46 & Avg Fwd Segment Size & 5.06E-05 \\
  47 & Total Fwd Packets & 4.99E-05 \\
  48 & Bwd Packets/s & 4.98E-05 \\
  49 & Idle Min & 4.57E-05 \\
  50 & Min Packet Length & 4.49E-05 \\
  51 & Flow IAT Min & 4.38E-05 \\

  \bottomrule
 \end{tabular}
 \label{tab:ganfs_ranking}
\end{table} %

\begin{table}[h!] %
 \caption{Features ranked according to the sensitivity by GAN Feature Selection Algorithm}
 \centering
 \small 
 \begin{tabular}{rlr}
  \toprule
  S.No. & Feature & Sensitivity\_Score \\
  \midrule
    52 & Bwd IAT Mean & 4.14E-05 \\
  53 & Packet Length Std & 3.85E-05 \\
  54 & Packet Length Mean & 3.78E-05 \\
  55 & Fwd Header Length.1 & 1.65E-05 \\
  56 & Fwd Header Length & 1.32E-05 \\
  57 & Flow IAT Max & 9.16E-06 \\
  58 & Destination Port & 6.09E-06 \\
  59 & Fwd IAT Max & 5.94E-06 \\
  60 & Source Port & 5.38E-06 \\
  61 & Flow Duration & 3.52E-06 \\
  62 & Fwd IAT Total & 3.11E-06 \\
  63 & Fwd IAT Mean & 1.66E-06 \\
  63 & Fwd IAT Mean & 1.66E-06 \\
  64 & Flow Bytes/s & 1.46E-06 \\
  65 & Flow IAT Mean & 1.16E-06 \\
  66 & Fwd Packets/s & 9.81E-07 \\
  67 & Flow Packets/s & 9.35E-07 \\
  68 & Fwd IAT Std & 7.57E-07 \\
  69 & Flow IAT Std & 4.86E-07 \\
  70 & Bwd PSH Flags & 0 \\
  71 & PSH Flag Count & 0 \\
  72 & Fwd URG Flags & 0 \\
  73 & FIN Flag Count & 0 \\
  74 & Fwd Avg Bytes/Bulk & 0 \\
  75 & ECE Flag Count & 0 \\
  76 & Bwd URG Flags & 0 \\
  77 & Bwd Avg Bulk Rate & 0 \\
  78 & Fwd Avg Packets/Bulk & 0 \\
  79 & Fwd Avg Bulk Rate & 0 \\
  80 & Bwd Avg Bytes/Bulk & 0 \\
  81 & Bwd Avg Packets/Bulk & 0 \\
  \bottomrule
 \end{tabular}
 \label{tab:ganfs_ranking_2}
\end{table} %

\end{document}